\documentclass[12pt]{article}
\usepackage{graphicx}
\begin{document}

\begin{center}
{\bf Midrapidity inclusive densities in high energy
$pp$ collisions in additive quark model}

\vspace{.2cm}

Yu.M. Shabelski and A.G. Shuvaev \\

\vspace{.5cm}

Petersburg Nuclear Physics Institute, Kurchatov National
Research Center\\
Gatchina, St. Petersburg 188300, Russia\\
\vskip 0.9 truecm
E-mail: shabelsk@thd.pnpi.spb.ru\\
E-mail: shuvaev@thd.pnpi.spb.ru

\vspace{1.2cm}

\end{center}

\begin{abstract}
\noindent
High energy (CERN SPS and LHC) inelastic $pp$ ($p\bar p$) scattering
is treated in the framework of additive quark model together with
Pomeron exchange theory. We extract the midrapidity inclusive density
of the charged secondaries produced in a single quark-quark collision
and investigate its energy dependence. Predictions for the $\pi p$
collisions are presented.
\end{abstract}


\section{Introduction}
Regge theory provides a useful tool for the phenomenological description
of high energy hadron collisions~\cite{Dr, RMK, MerShab, Sel}.
The quantitative predictions of Regge calculus are essentially dependent
on the assumed coupling of participating hadrons to the Pomeron.
In our previous papers~\cite{Shabelski:2014yba,Shabelski:2015bba} we described
elastic $pp$ ($p\bar p$) scattering and diffractive dissociation
processes including the recent LHC data
in terms of a simple Regge exchange approach in the framework
of the additive quark model (AQM)~\cite{LF,VH}, or constituent quarks
model as it is also referred to. It has been successfully applied
to $pp$ scattering processes at LHC energies~\cite{Bondarenko:2005pw}.
In the present paper we consider the inclusive densities of produced
secondaries in the midrapidity region in the same approach.

In AQM baryon is treated as a system of
three spatially separated compact objects -- the constituent quarks.
Each constituent quark is colored and has an internal quark-gluon
structure and a finite radius that is much less than the radius of
the proton, $r_q^2 \ll r_p^2$. This picture is in good agreement both with
$SU(3)$ symmetry of the strong interaction and the quark-gluon structure
of the proton~\cite{DDT, Shekhter, Anisovich}.
The constituent quarks play the roles of incident particles
in terms of which $pp$ scattering is described in AQM.

In the case of inelastic $pp$ collisions the secondary particles are
produced in AQM in one or several $qq$ collisions, so it
opens the possibility to investigate inclusive densities of
the secondaries
produced in the single $qq$ collision at the different initial
energies.
After that we can calculate the central inclusive densities
in $\pi p$ collisions without any new parameters.

\section{High energy $pp$ interactions in AQM}

Elastic amplitudes for large energy $s=(p_1+p_2)^2$
and small momentum transfer $t$ are dominated by Pomeron
exchange. We neglect the small difference in $pp$ and $p\bar p$
scattering coming from the exchange of negative signature Reggeons,
Odderon (see e.g.~\cite{Avila} and references therein),
$\omega$-Reggeon etc., since their contributions are suppressed by $s$.

\begin{figure}[htb]
\centering
\includegraphics[width=0.6\hsize]{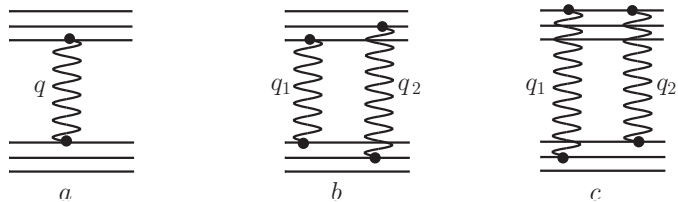}
\caption{\footnotesize The AQM diagrams for $pp$ elastic scattering. The
{\em straight lines} stand for quarks, the {\em wavy} lines denote Pomerons, $Q$ is the
momentum transferred, $t=-Q^2$. {\bf a} The one of the single Pomeron
diagrams, {\bf b} and~{\bf c} represent double Pomeron exchange with two
Pomeron coupled to the different quark~{\bf b} and to the same quarks~{\bf c},
$q_1+q_2=Q$.}
\label{1P2P}
\end{figure}

The single $t$-channel exchange results into
the amplitude of constituent quarks scattering,
\begin{equation}
\label{Mqq}
M_{qq}^{(1)}(s,t) = \gamma_{qq}(t) \cdot
\left(\frac{s}{s_0}\right)^{\alpha_P(t) - 1} \cdot
\eta_P(t) \;,
\end{equation}
where $\alpha_P(t) = \alpha_P(0) + \alpha^\prime_P\cdot t$
is the Pomeron trajectory specified by the intercept
and slope values $\alpha_P(0)$ and $\alpha^\prime_P$, respectively.
The Pomeron signature factor,
$$
\eta_P(t) \,=\, i \,-\, \tan^{-1}
\left(\frac{\pi \alpha_P(t)}2\right),
$$
determines the complex structure of the amplitude. The factor
$\gamma_{qq}(t)=g_1(t)\cdot g_2(t)$ has the meaning
of the Pomeron coupling to the beam and target particles,
the functions $g_{1,2}(t)$ being the vertices of the constituent
quark-Pomeron interaction (filled circles in Fig.~\ref{1P2P}).
It is worth to emphasize that the $qq$ interaction is described
here by single effective Pomeron exchange between each
$qq$ pair. Generally it may include the contributions
of several Gribov bare Pomerons \cite{Gribov}
and the parameters of the effective Pomeron could
be different from those of the bare Pomerons.
At the same time the one quark interaction
with the two different target quarks is mediated
by the exchange of the two effective Pomerons.
In this respect there is a close resemblance between
the nucleon-nucleon scattering in AQM and the nucleus-nucleus
scattering in Glauber theory. The $qq$ interaction plays
the same role in the first case as the $NN$ interaction
in the second.

The elastic $pp$ scattering amplitude
(or $p\bar p$ scattering amplitude, here we do not distinguish between
the two) is basically expressed as
\begin{equation}
\label{VQQ}
M_{pp}(s,t)\,=\,\int dK\,dK^\prime
\psi^*(k_i^\prime+Q_i^{\,\prime})\,\psi^*(k_i+Q_i)\,
V(Q,Q^{\,\prime})\,\psi(k_i^\prime)\,\psi(k_i).
\end{equation}
In this formula $\psi(k_i)\equiv \psi(k_1,k_2,k_3)$,
is the initial proton wave function in terms of the quarks'
transverse momenta $k_i$, while
$\psi(k_i+Q_i)\equiv \psi(k_1+Q_1,k_2+Q_2,k_3+Q_3)$
is the wavefunction of the scattered proton.
The interaction vertex $V(Q,Q^{\,\prime})\equiv
V(Q_1,Q_2,Q_3,Q_1^{\,\prime},Q_2^{\,\prime},Q_3^{\,\prime})$
stands for the multipomeron exchange, $Q_k$ and $Q_l^{\,\prime}$
are the momenta transferred to the target quark $k$ or beam
quark $l$ by the Pomerons attached to them,
$Q$ is the total momentum transferred
in the scattering, $Q^2=Q^{\prime\,2}=-t$.

The scattering amplitude is presented in AQM as a sum over
the terms with a given number of Pomerons,
\begin{equation}
\label{totamp}
M_{pp}(s,t)\,=\,\sum_n M_{pp}^{(n)}(s,t),
\end{equation}
where the amplitudes $M_{pp}^{(n)}$ collect all
diagrams comprising various connections of the beam and target
quark lines with $n$ Pomerons.
Similar to Glauber theory \cite{Glaub, FG} one has to rule out
the multiple interactions between the same quark pair.
AQM permits the Pomeron to connect any two quark lines only once.
It crucially decreases the combinatorics, leaving the diagrams
with no more than $n=9$ effective Pomerons.
Several AQM diagrams are shown in Fig.~\ref{1P2P}.

In the following we assume the Pomeron trajectory
to have the simplest form,
$$
\left(\frac{s}{s_0}\right)^{\alpha_P(t) - 1}\,=\,e^{\Delta\cdot\xi}
e^{-r_q^2\,q^2}, ~~ \xi\equiv \ln\frac{s}{s_0},~~
r_q^2\equiv \alpha^\prime\cdot\xi.
$$
The value $r_q^2$ defines the radius of the quark-quark interaction,
while $S_0=(9~{\rm GeV})^2$ has the meaning of typical energy scale
in Regge theory.

In the first order there are nine equal quark-quark contributions
due to one Pomeron exchange between $qq$ pairs. The amplitude
(\ref{VQQ}) reduces to a single term with $Q_1=Q_1^{\,\prime} = Q$,
$Q_{2,3}=Q_{2,3}^{\,\prime} = 0$,
\begin{equation}
\label{M1}
M_{pp}^{(1)}\,=\,9\biggl(\gamma_{qq}\eta_P(t) e^{\Delta\cdot\xi}
\biggr)\,e^{-r_q^2\,Q^2}F_P(Q,0,0)^2,
\end{equation}
expressed through the overlap function
\begin{equation}
\label{FP}
F_P(Q_1,Q_2,Q_3)\,=\,\int dK\,\psi^*(k_1,k_2,k_3)\,
\psi(k_1+Q_1,k_2+Q_2,k_3+Q_3).
\end{equation}
The function $F_P(Q,0,0)$ plays the role of a proton form factor
for the strong interaction in AQM.

The quarks' wave function has been taken in the simple form
of Gaussian packets,
\begin{equation}
\label{gausspack}
\psi(k_1,k_2,k_3)\,=\,N\bigl[\,e^{-a_1(k_1^2+k_2^2+k_3^2)}\,
+\,C_1\,e^{-a_2(k_1^2+k_2^2+k_3^2)}
+\,C_2\,e^{-a_3(k_1^2+k_2^2+k_3^2)}\bigr],
\end{equation}
normalized to unity.
The parametrization by the single exponent
is unable to reproduce the minimum in $d\sigma/dt$ distribution
evidently seen in the experimental data for $\sqrt{s}=7$~TeV
\cite{TO1,TO2}.
The two exponential fit used in our previous
papers~\cite{Shabelski:2014yba,Shabelski:2015bba}
reproduces this minimum but gives too low values of $d\sigma/dt$
at $|t| \sim 0.7 - 0.8$~Gev$^2$. In the present paper the wave
function is parameterized by the sum of three Gaussian exponents,
which allows for the better description of $d\sigma/dt$.

Now the parameters read
$$
\Delta=0.14,~~~
\alpha^\prime=0.116\,{\rm GeV}^{-2},~~~
\gamma_{qq}=0.45\,{\rm GeV}^{-2}.
$$
$$
a_1=9.0\,{\rm GeV}^{-2},~~~
a_2=0.29\,{\rm GeV}^{-2},~~~
a_3=2.0\,{\rm GeV}^{-2},~~~
C_1=0.024,~~~
C_2=0.05.
$$
One has to remark here that we do not claim the real
matter distribution inside the proton to be close
to the Gaussian shape, this form is suitable
only to perform all the integrals analytically.
The value $a_1$ is quite compatible with the large proton size assumed
above, whereas $a_{2,3}$ values manifest the presence of the small radius
components in the proton wave function. However, their relative weights
are small so the total wave function (\ref{gausspack})
matches the condition $r_p^2 \gg r_q^2$, which is effectively
fulfilled for the mean radii that are important
for the calculations validity.

The higher orders elastic terms are expressed through
the functions (\ref{FP}) integrated over Pomerons' momenta,
\begin{eqnarray}
\label{Mn}
M_{pp}^{(n)}(s,t)\,&=&\,i^{n-1}\biggl(\gamma_{qq}\eta_P(t_n)
e^{\Delta\cdot\xi}\biggr)^n\,
\int\frac{d^2q_1}{\pi}\cdots \frac{d^2q_n}{\pi}
\,\pi\,\delta^{(2)}(q_1+\ldots +\,q_n-Q)\,\\
&&\times\,e^{-r_q^2(q_1^2+\ldots + q_n^2)}\,
\frac 1{n!}\sum\limits_{n~\rm connections}\hspace{-1.5em}
F_P(Q_1,Q_2,Q_3)\,F_P(Q_1^{\,\prime},Q_2^{\,\prime},Q_3^{\,\prime}),
~~~t_n\simeq t/n. \nonumber
\end{eqnarray}
The sum in this formula refers to all distinct ways
to connect the beam and target quark lines with $n$ Pomerons
in the scattering diagram. The set of momenta $Q_i$ and $Q_l^{\,\prime}$
the quarks acquire from the attached Pomerons is particular
for each connection pattern. A more detailed description can be found
in~\cite{Shabelski:2014yba}.

With the amplitude (\ref{totamp}) the differential cross section
in the normalization adopted here is evaluated as
\begin{equation}
\label{ds/dt}
\frac{d\sigma}{dt}\,=\,4\pi\,\bigl|M_{pp}(s,t)\bigr|^2\,
=\,4\pi\,\bigl[\bigl({\rm Re}\, M_{pp}(s,t)\bigr)^2
+ \bigl({\rm Im}\, M_{pp}(s,t)\bigr)^2\bigr].
\end{equation}
The optical theorem, which relates the total cross section
and the imaginary part of the amplitude, in this normalization reads
\begin{equation}
\label{Opth}
\sigma_{pp}^{tot}\,=\,8\pi\,{\rm Im}\, M_{pp}(s,t=0).
\end{equation}

The condition for the AQM applicability,
$r_q^2/r_p^2 \ll 1$, holds rather well since
$r_p^2\simeq 12$~Gev$^{-2}$
whereas $r_q^2\simeq 1.5$~Gev$^{-2}$ at
$\sqrt s\approx 7$~TeV.

The resulting differential cross sections for $pp$ scattering
at $\sqrt s = 7$~TeV are presented in Fig.~\ref{pp}
together with the predictions for $\pi p$
scattering at the same energy (see below).
\begin{figure}[htb]
\centering
\vskip -1.cm
\includegraphics[width=.45\textwidth]{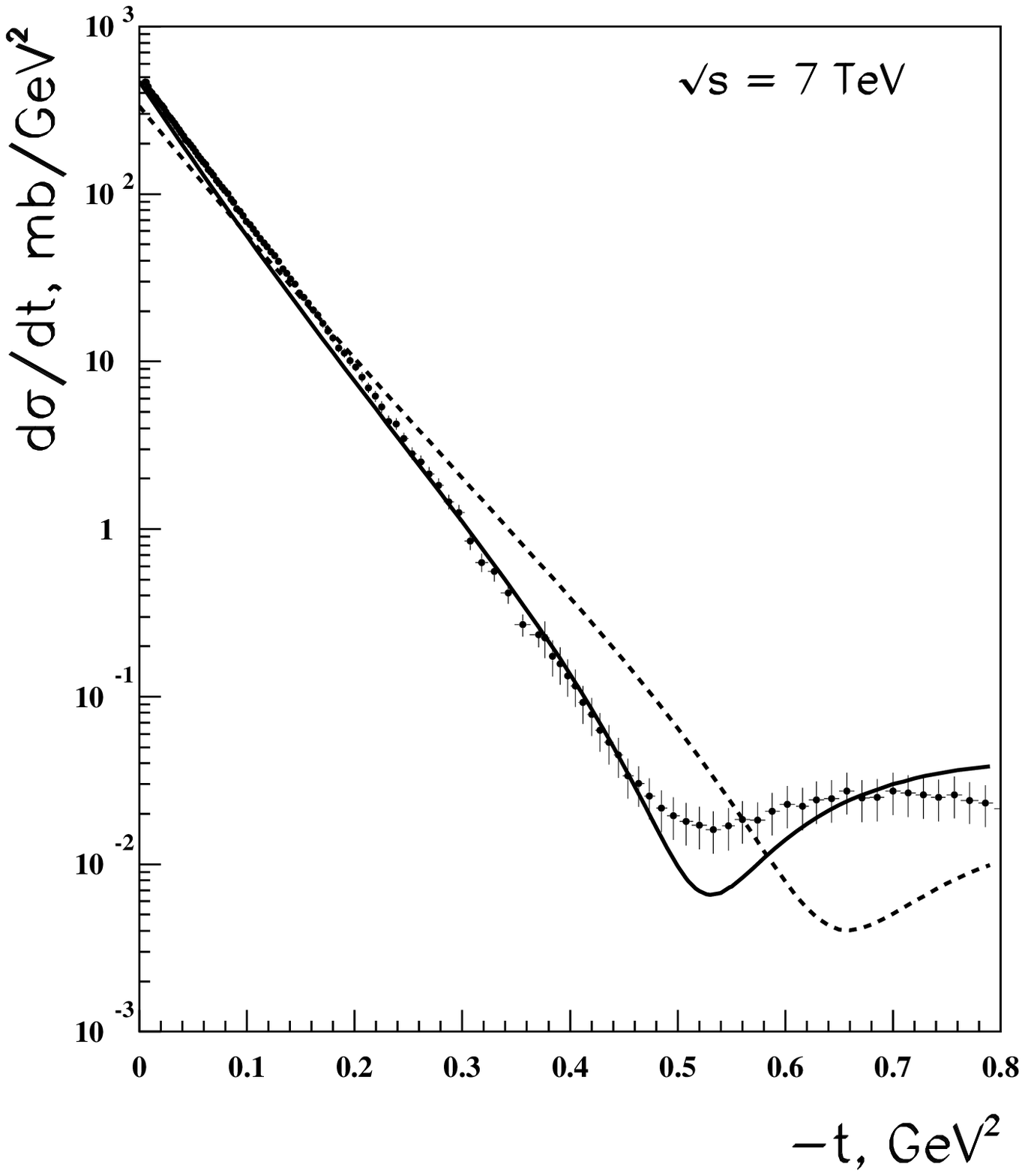}
\includegraphics[width=.45\textwidth]{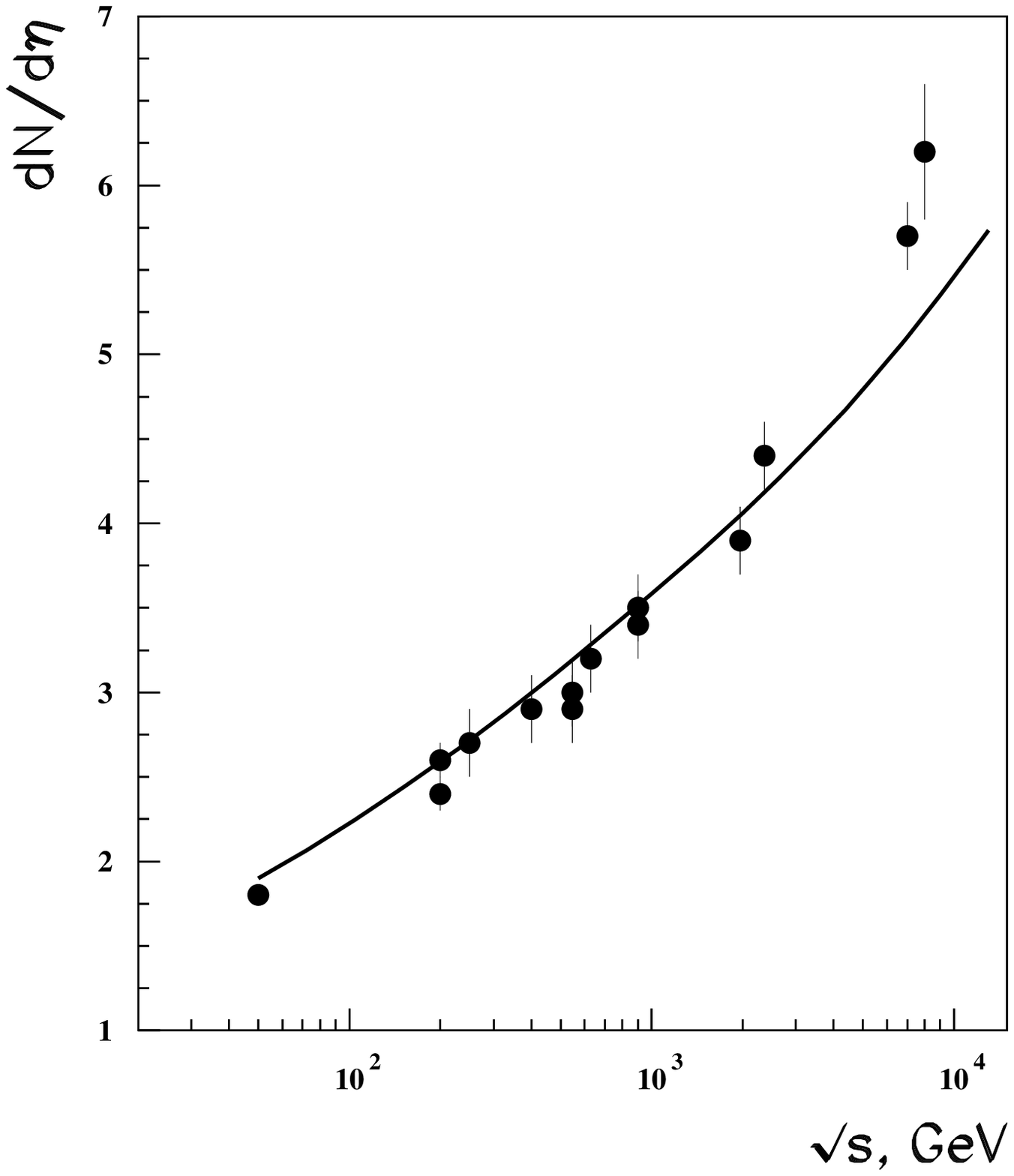}
\caption{\footnotesize {\em Left} Differential cross section
of the elastic $pp$ scattering
({\em solid line}) and $\pi p$ scattering ({\em dashed line}),
see Sect.4,
at $\sqrt s = 7$~TeV. The experimental points
for $pp$ scattering have been taken from~\cite{TO1,TO2}.
{\em Right} Pseudorapidity distribution of the secondaries
$dN^{\rm NSD}/d\eta$ in $pp$ scattering for
the non-single diffractive events.
The {\em solid line} shows the AQM estimates by
Eq.~(\ref{NSD}).
The experimental points have been taken
from~\cite{Chatrchyan:2014qka}}
\label{pp}
\end{figure}

\section{AGK cuts and inclusive densities in $pp$ and $qq$
interactions}

\begin{figure}[htb]
\centering
\includegraphics[width=0.6\hsize]{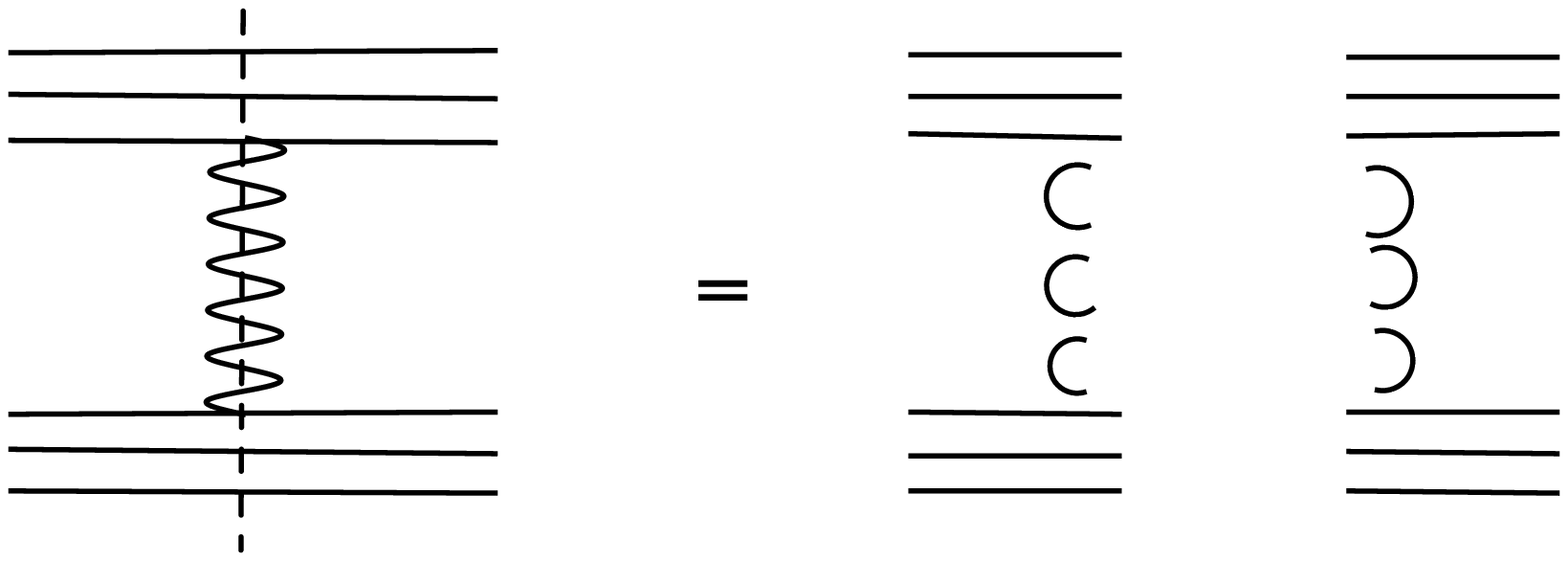}
\caption{\footnotesize The first quark order diagram
contributing to the inelastic particle production in
$pp$ collision}
\label{1Pcut}
\end{figure}

All amplitudes of the inelastic processes in high energy
$pp$ collisions can be treated as the sum of various
absorptive parts of elastic $pp$ amplitude;
see AGK cutting rules~\cite{AGK}.
In the AQM the diagram with a single $qq$ interaction,
Fig.~\ref{1Pcut}, has only one absorptive part.
The one-Pomeron cut in the left hand side of Fig.~\ref{1Pcut}
corresponds to the multiperipheral ladder of the produced
secondaries in the right hand
side of Fig.~\ref{1Pcut}. The resulting
cross section is $\sigma^{(1)}$.

\begin{figure}[htb]
\includegraphics[width=.45\textwidth]{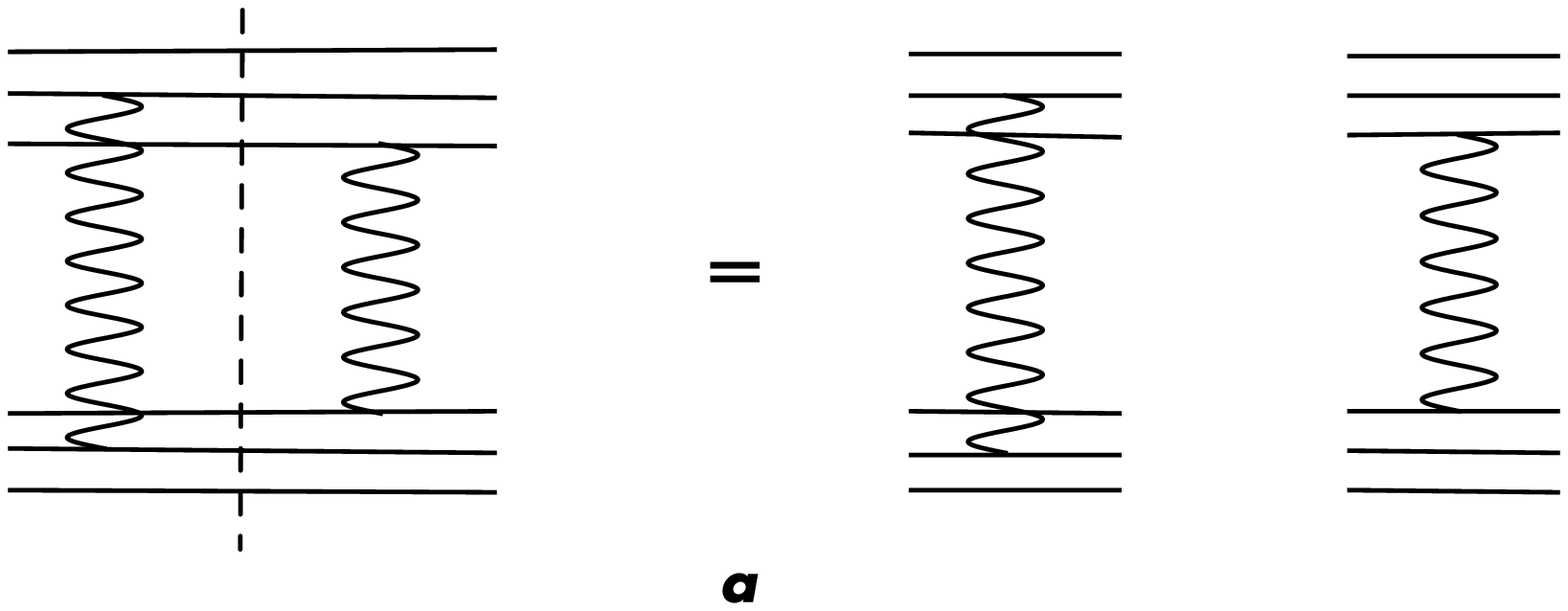}
\hskip 1cm
\includegraphics[width=.43\textwidth]{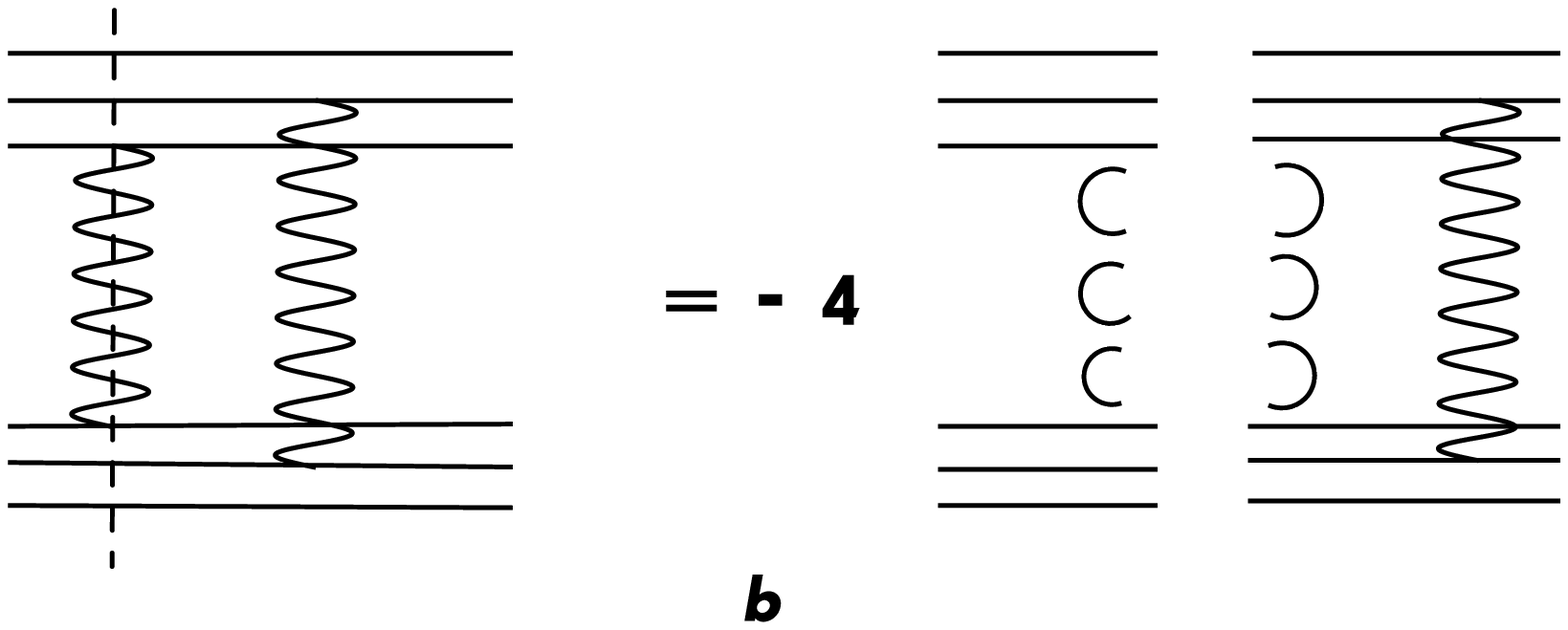}
\centering
\includegraphics[width=.4\textwidth]{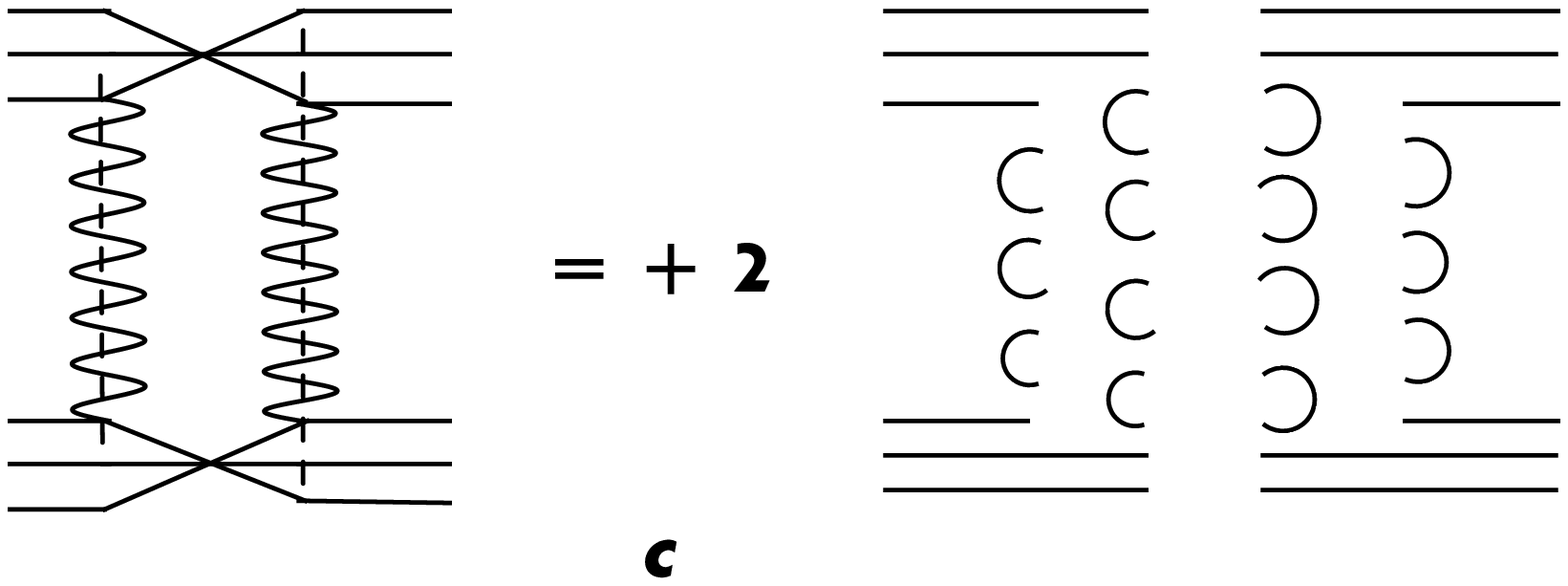}
\caption{\footnotesize The diagrams with a double interaction}
\label{2Pcut}
\end{figure}

In the case of double interaction in Fig.~\ref{1P2P}b
the imaginary part is given by the sum of three different
absorptive parts presented
in Fig.~\ref{2Pcut}. The first one,
the cut between Pomerons, is shown in Fig.~\ref{2Pcut}a.
It describes the elastic
and diffractive dissociation processes without production
of secondaries in the central (midrapidity) region.
The second absorptive part, Fig.~\ref{2Pcut}b, corresponds
to the cut of one Pomeron and gives the first rescattering
correction to the processes
in Fig.~\ref{1Pcut}.
The multipheripheral ladders in Figs.~\ref{1Pcut}
and \ref{2Pcut}b
are practically the same and have the same midrapidity inclusive
densities $dN_{qq}/dy$.
The absorptive part in Fig.~\ref{2Pcut}b
has the numerical factor -4 due to the combinatorics \cite{AGK}.
The third absorptive part is shown in Fig.~\ref{2Pcut}c,
where the cut slices
both Pomerons. It means the simultaneous production
of two multipheripheral ladders
of the secondaries. These ladders are also practically
identical to those
in Figs.~\ref{1Pcut} and \ref{2Pcut}b and result
in equal inclusive densities $dN_{qq}/dy$
(we neglect the very small numerical difference coming from
the energy conservation). The combinatorial factor here is
+2 \cite{AGK}.
The contribution to the inclusive density of the secondaries
from Fig.~\ref{2Pcut}c
is $4dN_{qq}/dy$. It is compensated by the negative
contribution $-4dN_{qq}/dy$
from the process in Fig.~\ref{2Pcut}b. Finally,
the sum of all absorptive
parts collected in Fig.~\ref{2Pcut}
yields zero contribution to the inclusive densities
of secondary particles
in complete agreement with the AGK cutting rules~\cite{AGK}.

Similarly, there is no contribution to the resulting
inclusive density due to diagrams with a larger number
of quark-quark interaction, therefore it is only
the impulse approximation diagrams in Fig.~\ref{1Pcut}
that provide the inclusive density of the secondaries
produced in $pp$ collisions in the midrapidity region,
\begin{equation}
\label{Eqy}
\sigma_{pp}^{inel}\cdot \frac{dN_{pp}}{dy}
= 9\,\sigma_{qq}^{(1)}\cdot \frac{dN_{qq}}{dy}.
\end{equation}
where $\sigma_{qq}^{(1)}$ is the first order contribution
to the total cross section.
This equation is true as well for the pseudorapidity
inclusive densities, which amounts to the replacement
$dy \to d\eta$.

The cross section of $pp$ interaction used in~(\ref{Eqy})
depends on the way
the value of the inclusive density is fixed.
It is determined as the number of the secondaries divided
by the number of events
in the small interval $dy$ in the midrapidity region.
In the diffractive dissociation
the secondaries are produced practically only
in the fragmentation regions; therefore,
the number of secondary particles in the midrapidity
region does not change
whether or not we include diffractive
dissociation events in our sample.
However, the number of events, i.e. the denominator
in the definition of $dN/dy$,
differs for these two cases, so the net value $dN_{pp}/dy$
in the left hand side of (\ref{Eqy}) has to be multiplied
by $\sigma_{PP}^{inel}$,
if we take all inelastic events,
or by $\sigma_{PP}^{nondiffr}$
if we take the events without diffractive dissociation.

\begin{figure}[htb]
\centering
\includegraphics[width=0.45\textwidth]{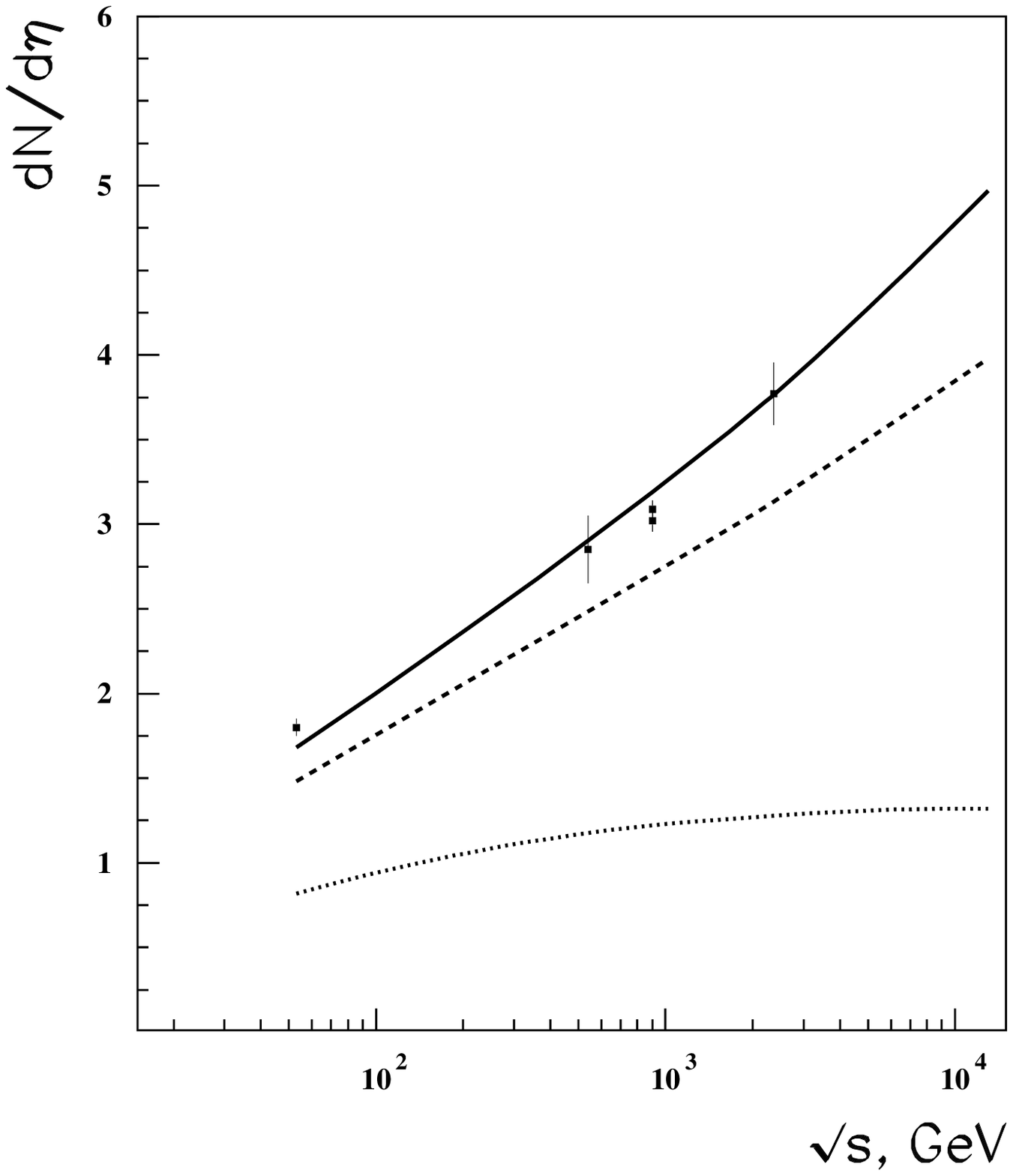}
\includegraphics[width=0.45\textwidth]{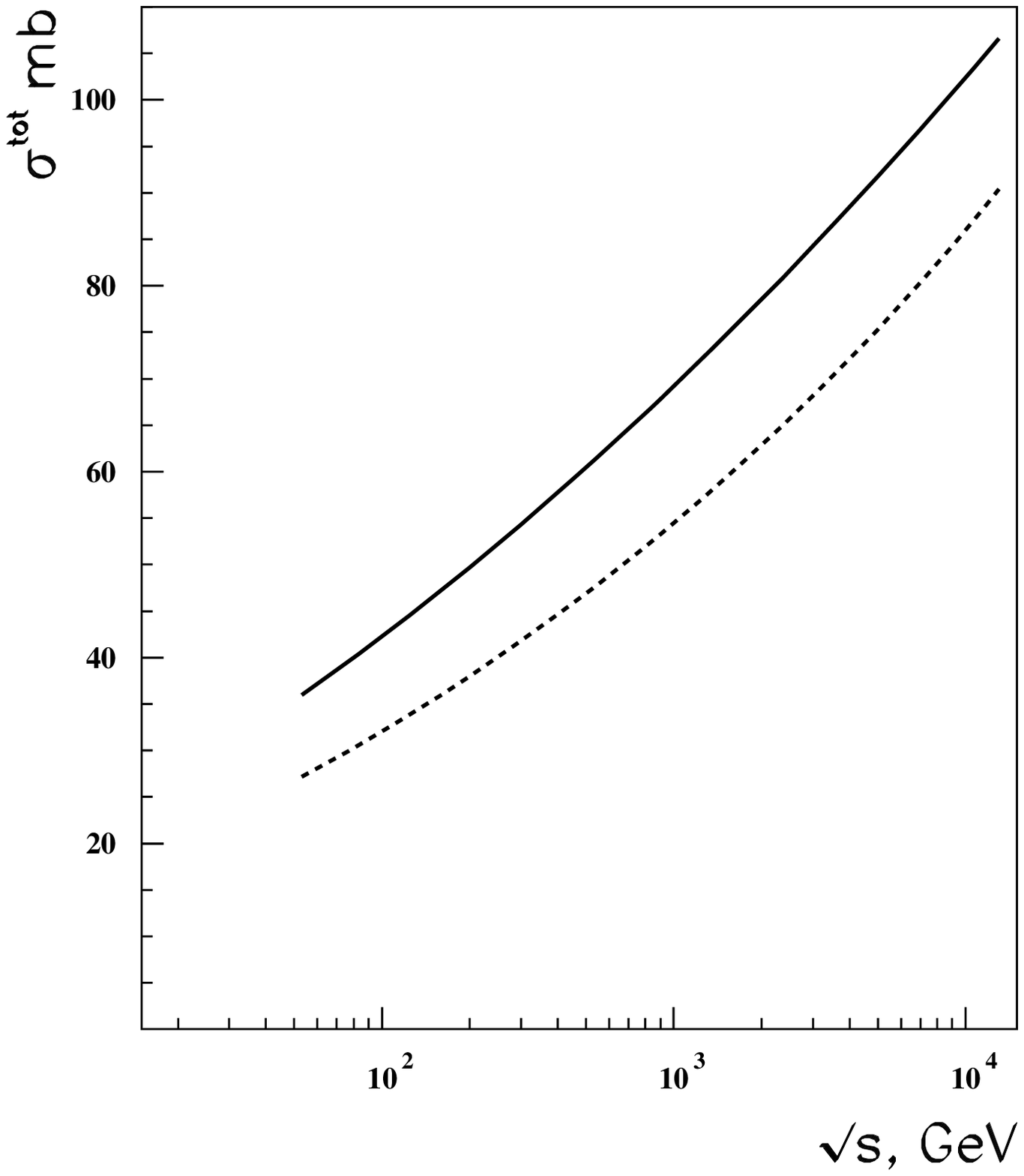}
\vskip -2cm
\caption{\footnotesize
{\em Left} The pseudorapidity distributions
of all charged secondaries
produced in the inelastic $pp$ and $p\bar p$ collisions
at different energies
together with their description in QGSM ({\em solid curve}).
The {\em dotted line} shows the extracted $dN_{qq}/d\eta$ values.
The {\em dashed line} presents the obtained pseudorapidity distribution
of charged secondaries in the inelastic $\pi p$ collision.
The experimental points are taken from
refs.~\cite{Kaidalov, Aamodt:2010ft}.
{\em Right} The total cross sections of $pp$ ({\em solid line}) and $\pi p$
({\em dashed line}) as the initial energy functions.
The experimental $pp$ points are taken from
\cite{Am, TOT1a}}
\label{secondary}
\end{figure}

Let us try to estimate the energy dependence
of $dN_{qq}/dy$ and $dN_{qq}/d\eta$
using the existing data.
There are several available experimental points
for $dN_{pp}/d\eta$ at the different energies measured
in all inelastic events.
They are shown in Fig.~\ref{secondary} together with
the calculations of $dN_{pp}/dy$ and $dN_{pp}/d\eta$ in
the quark-gluon string model (QGSM)~\cite{Merino:2011sq, Merino:2012wv}.
Actually QGSM output is employed here only to extrapolate
the existing experimental data.

The inelastic cross section entering Eq.~(\ref{Eqy})
can be obtained from the identity
\begin{equation}
\label{sinel}
\sigma_{pp}^{inel}\,=\,\sigma_{pp}^{tot} - \sigma_{pp}^{elastic},
\end{equation}
where the total cross section is evaluated through
the optical theorem (\ref{Opth}) by summing up all nine orders
of AQM diagrams~(\ref{Mn}), while the elastic cross section,
$\sigma_{pp}^{elastic}$, is obtained
by integrating differential cross section (\ref{ds/dt})
over $t$.
The cross section $\sigma_{qq}^{(1)}$
is given by the first order of the AQM amplitude (\ref{M1}).

Equations (\ref{Eqy}) and (\ref{sinel})
allow one to find $dN_{qq}/d\eta$ values presented
in Fig.~\ref{secondary}.
At the LHC energies $\sqrt s > 0.9$~TeV $dN_{qq}/d\eta$ becomes
independent on the initial energy
within our theoretical accuracy $\sim 10\%$.

Experimental papers often present the data for the energy dependence
of the pseudorapidity distribution of the secondaries,
$dN^{\rm NSD}/d\eta$, measured in the non-single diffractive events.
It can be obtained by the formula
\begin{equation}
\label{NSD}
\frac{dN^{\rm NSD}}{d\eta}\,=\,\frac{dN}{d\eta}\,
\frac{\sigma_{pp}^{inel}}{\sigma_{pp}^{inel}-2\sigma_{pp}^{\rm SD}},
\end{equation}
where $\sigma_{pp}^{\rm SD}$ is the cross section of the single
diffractive $pp$ scattering (from one side). To make a quick estimate
we have used $\sigma_{pp}^{\rm SD}$ values calculated in AQM in our
previous paper~\cite{Shabelski:2015bba}. The results are shown in
Fig.~\ref{pp} (right panel) together with the existing experimental points.
We get a reasonable agreement, the ratio
$\frac{dN^{\rm NSD}}{d\eta}/\frac{dN}{d\eta}\sim 1.1\div 1.15$
at the LHC energies.

\section{Predictions for $\pi p$ collisions}

To obtain the predictions of midrapidity inclusive densities in
$\pi p$ collisions one needs to know the total $\pi p$ cross
section. It has not been measured experimentally
at the very high energies
but can be calculated in AQM.
In our approach the interaction of quarks and
antiquarks constituting the pion are the same as those in the proton
(so far as only Pomeron exchange is encountered). The amplitude of
the elastic $\pi p$ collision is evaluated in AQM similarly to
the elastic $pp$ one, see (\ref{Mn}),
\begin{eqnarray}
\label{Mpip}
M_{\pi p}^{(n)}(s,t)\,&=&\,i^{n-1}\biggl(\gamma_{qq}\eta_P(t_n)
e^{\Delta\cdot\xi}\biggr)^n\,
\int\frac{d^2q_1}{\pi}\cdots \frac{d^2q_n}{\pi}
\,\pi\,\delta^{(2)}(q_1+\ldots +\,q_n-Q)\,\\
&&\times\,e^{-r_q^2(q_1^2+\ldots + q_n^2)}\,
\frac 1{n!}\sum\limits_{n~\rm connections}\hspace{-1.5em}
F_\pi(Q_1,Q_2)\,F_P(Q_1^{\,\prime},Q_2^{\,\prime},Q_3^{\,\prime}),
~~~t_n\simeq t/n. \nonumber
\end{eqnarray}
Here $F_\pi$ and $F_P$ are the pion and proton form factors
while all other variables are the same as those for $pp$
scattering.

\begin{figure}[htb]
\centering
\includegraphics[width=.6\textwidth]{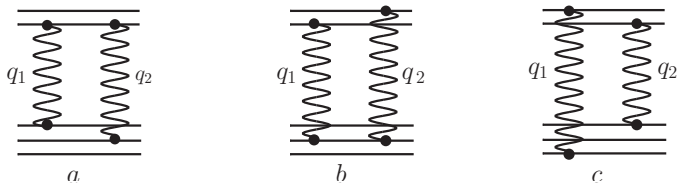}
\caption{\footnotesize Second order AQM diagrams
for $\pi p$ scattering.}
\label{pip aqm}
\end{figure}

The quark combinatorics is more simple for $\pi p$ collisions
compared to $pp$ case. In particular, there are only six orders of the
admissible diagrams. The sum for the first order contribution
reduces to a single term, $6F_\pi(Q,0)\,F_P(Q,0,0)$, $Q^2=-t$. The
second order sum includes three types of diagrams,
\begin{eqnarray}
&&\frac 1{2!}\,\sum\limits_{2~\rm connections}
F_\pi(Q_1,Q_2)\,F_P(Q_1^{\,\prime},Q_2^{\,\prime},Q_3^{\,\prime})
\nonumber \\
&&\,=\,
6\,F_\pi(Q,0)F_P(q_1,q_2,0)\,+\,3\,F_\pi(q_1,q_2)F_P(Q,0,0)
+\,6\,F_\pi(q_1,q_2)F_P(q_1,q_2,0),\nonumber\\
&&~~~~Q\,=\,q_1\,+\,q_2,
\nonumber
\end{eqnarray}
where the first two terms come from the diagram
with both Pomerons coupled to the same quark line
in the pion (first term, Fig.~\ref{pip aqm}a) and in the proton
(second term, Fig.~\ref{pip aqm}b); in the third term they
connect different quark lines (Fig.~\ref{pip aqm}c).
The numerical coefficients encounter the number of connections
resulting in equal expressions after variables changing in
the integrals~(\ref{Mpip}).

The rest orders have a similar
structure derived from the combinatorics
to redistribute $q_1,\ldots, q_n$ momenta among $Q_i$
and $Q_i^{\,\prime}$ groups.
In the highest order, containing six effective Pomerons,
$$
\frac 1{6!}\sum\limits_{6~\rm connections}
F_\pi(Q_1,Q_2,Q_3)\,F_P(Q_1^{\,\prime},Q_2^{\,\prime},
Q_3^{\,\prime})
$$
$$
=\,F_\pi(q_1 + q_2 + q_3,q_4 + q_5 + q_6)
F_P(q_1 + q_4, q_2+q_5,q_3 + q_6),
$$
each quark from the proton interacts with both the quarks from
the pion.

The differential and the total $\pi p$ cross sections
are evaluated via Eq.~(\ref{ds/dt})
and the optical theorem (\ref{Opth}) respectively.
Our results for the elastic $\pi p$ scattering show a minimum
at $\sqrt s = 7$~TeV placed at $-t\approx 0.65$~GeV$^2$
(Fig~\ref{pp}).
The ratio of the total $pp$ and $\pi p$ cross section in the optical
approximation of AQM is well known to be 3/2 \cite{LF}.
With the multiple rescattering included this value changes
depending on the ratio of proton and pion radii.
The experimental data \cite{Bernard:2000qz} gives for
the ratio $r_\pi^2/r_p^2\approx 0.57$,
so we take the pion wave function in the same form,
(\ref{gausspack}), rescaling all radius parameters
as $a_{1,2,3}^\pi = 0.57 a_{1,2,3}^p$.
Actually the dependence $\sigma_{\pi p}^{tot}$ on the parameters
of $a_{1,2,3}$ and $C_{1,2}$
is rather weak. We get the ratio
$\sigma_{pp}^{tot}/\sigma_{\pi p}^{tot} \approx 1.2 \div 1.3$.
Unfortunately there are no experimental data on the $\pi p$
scattering at the LHC energies, the AQM results for them
are presented in Fig.~\ref{secondary} together with the predictions
for the $pp$ case. Note here as well that AQM predictions
for $d\sigma_{pp}/dt(t=0)$ are in good agreement with
the data~\cite{Shabelski:2014yba}.

The obtained values $\sigma_{\pi p}^{tot}$ allow to find
the midrapidity inclusive density in $\pi p$ collisions.
The results for $dN_{\pi p}/d\eta(\eta=0)$ as a function of
the initial energy are presented in Fig.~\ref{secondary}.
The obtained data can be used for the calculation of particle production
at the very high energies; in particular, in cosmic ray physics.

\section{Conclusion}

In the framework of AQM we have extracted the inclusive density
of the secondaries in $qq$ interactions in the midrapidity region.
We used these values to get prediction for $\pi p$ collisions
at high energies.
These quantities can be useful to estimate the secondary
production at the very high energies, say, in cosmic ray physics.

The applicability of AQM requires the contribution from
the multipomeron $qq$ interactions
to be small
compared to the interaction between different quarks responsible
in this approach for the $pp$ scattering. This is valid for
the soft processes, whose amplitude is practically pure imaginary
so that the $qq$ cross section does not exceed the geometrical limit
$\sim r_q^2$. On the other hand there are additional combinatorial
factors increasing the $pp$ cross section, so it can always be assumed
to be larger than the $qq$ one. A reasonable description
of the elastic $pp$ scattering has been reached without appealing
to the enhanced diagrams with interacted Pomerons.
It provides evidence that AQM
is at work up to LHC energies.
However, for the energies essentially above the LHC ones
the multipomeron interactions would begin
to play an important role, which could modify our results for
asymptotically high energies.

This work has been supported by RSCF grant No 14 - 22 - 00281.

\end{document}